\documentclass[secnumarabic,amssymb, nobibnotes, aps, prd,superscriptaddress]{revtex4-1}

\usepackage{graphicx}
\usepackage{dcolumn}
\usepackage{bm}

\usepackage{amsmath,amssymb,amsfonts}
\usepackage{fancyhdr}
\usepackage{lipsum}
\usepackage{subcaption}
\usepackage{multirow}
\usepackage{tikz,comment}

\begin{document}

\title{Speech can produce jet-like transport relevant to asymptomatic spreading of virus}


\author{Manouk Abkarian}%
\thanks{M. A. and S. M. contributed equally.}
\affiliation{Centre de Biochimie Structurale, CNRS UMR 5048—INSERM UMR 1054, University of Montpellier, 34090 Montpellier, France}
\author{Simon Mendez}%
\thanks{M. A. and S. M. contributed equally.}
\affiliation{Institut Montpelli\'{e}rain Alexander Grothendieck, CNRS, University of Montpellier, 34095 Montpellier, France}
\author{Nan Xue}%
\affiliation{Department of Mechanical and Aerospace Engineering, Princeton University, Princeton, NJ 08544 USA}
\author{Fan Yang}%
\affiliation{Department of Mechanical and Aerospace Engineering, Princeton University, Princeton, NJ 08544 USA}
\author{Howard A. Stone}%
\thanks{To whom correspondence should be addressed. E-mail: hastone@princeton.edu}
\affiliation{Department of Mechanical and Aerospace Engineering, Princeton University, Princeton, NJ 08544 USA}

\begin{abstract}
Many scientific reports document that asymptomatic and presymptomatic individuals contribute to the spread of COVID-19, probably during conversations in social interactions. Droplet emission occurs during speech, yet few studies document the flow to provide the transport mechanism. This lack of understanding prevents informed public health guidance for risk reduction and mitigation strategies, e.g. the ``six-foot rule''. Here we analyze flows during breathing and speaking, including phonetic features, using order-of-magnitudes estimates, numerical simulations, and laboratory experiments. We document the spatio-temporal structure of the expelled air flow.  Phonetic characteristics of plosive sounds like `P' lead to enhanced directed transport, including jet-like flows that entrain the surrounding air. We highlight three distinct temporal scaling laws for the transport distance of exhaled material including (i) transport over a short distance ($<$ 0.5 m) in a fraction of a second, with large angular variations due to the complexity of speech,  (ii) a longer distance, approximately 1 m, where directed transport is driven by individual vortical puffs corresponding to plosive sounds, and (iii) a distance out to about 2 m, or even further, where sequential plosives in a sentence, corresponding effectively to a train of puffs, create conical, jet-like flows. The latter dictates the long-time transport in a conversation. We believe that this work will inform thinking about the role of ventilation, aerosol transport in disease transmission for humans and other animals, and yield a better understanding of linguistic aerodynamics, i.e., aerophonetics.

\end{abstract}

\maketitle

\section*{Introduction}
The rapid spread of COVID-19, the disease caused by the virus SARS-CoV-2, highlights the lack of guidelines and mitigation strategies for reducing the impact of airborne viruses in the absence of a vaccine. The inherent structural features of the air flows created by exhalation and inhalation during speech or simple breathing could be a potent yet, until recently, unsuspected transport mechanism for pathogen  transmission. This important topic surrounding viral transmission has largely been absent from the fluid mechanics and transport phenomena literature, and even absent more generally from quantitative studies of virus transport in the public health realm. We take steps toward quantifying fluid dynamic characteristics of this transmission pathway, which in the case of COVID-19, has been suggested to be associated with asymptomatic and presymptomatic carriers during relatively close social interactions, like breathing, speaking, laughing and singing. We focus on identifying and quantifying the complex flows associated with breathing and speaking; important areas for future research are indicated also. We recognize that much remains to be done, including integrating the findings and ideas here with potential mitigation strategies. 

There are many recent news articles reporting on the possibility of virus transmission during everyday social interactions. For example, documented cases include parties at homes, lunches at restaurants~\cite{RestaurantInfection}, side-by-side work in relatively confined spaces~\cite{ParkKoreanCallCenter}, choir practice in a small room~\cite{ChoirPractice}, fitness classes~\cite{FitnessClass}, a small number people  in a face-to-face meeting \cite{Hijnen2020}, etc.
Also, an editorial in the New England Journal of Medicine summarizes  differences between SARS-CoV-1, which is primarily transmitted from symptomatic individuals by respiratory droplets after virus replication in the lower respiratory tract, and SARS-CoV-2, for which viral replication and shed apparently occur most in the upper respiratory tract and do so even for asymptomatic individuals~\cite{GandhiYokoeHavlir}. These differences were suggested to be at least one reason why public health measures that were successful for SARS-CoV-1 have been much less effective for SARS-CoV-2.

Much has been written over many decades about  droplet shedding and transport during sneezing and coughing \cite{Wells1934,duguid1946size,Johnson2011,BourouibaBush2014,MittalNiSeo2020}. There remain open questions about the long-range transport of droplet nuclei or aerosols resulting from droplet evaporation  \cite{Bourouiba2020}, which is important to understand virus transmission from symptomatic individuals in all airborne respiratory diseases. 
In addition, researchers in the last decades have shown that droplet emission also occurs during speech \cite{duguid1946size,Johnson2011,AsadiRistenpart2019,AsadiRistenpart2020}, yet there are few quantitative studies of the corresponding breathing and speaking flows that provide the transport mechanism for such aerosols. For example, experiments and numerical simulations, based on scale models involving mannequins in rooms, have been used to study droplet transport and potential infection risk, e.g. ~\cite{NielsenOlmedoEtAl,LiuLiNielsenEtAl,AiMelikovReview}, including large-scale flow visualization studies of model out-flows~\cite{OzcanMeyerMelikov,XuNielsenEtAl,FengYaoEtAL} and the influence of ventilation strategies~\cite{OlmedoNielsenEtAl}. 

In this paper, we take first steps towards characterizing the fluid dynamics of speech. 
For example, questions that motivate our paper include how does an asymptomatic or a presymptomatic individual affect their surroundings by breathing, speaking, laughing or singing? What are the corresponding spatio-temporal features that quantify these changes and how do they affect the transport of exhaled material? 
Is there a better position or orientation to adopt when in a social interaction at a cafe, party, or workplace to minimize potential risk associated with the exhaled air from a speaker nearby?

We will illustrate that  there is a characteristic, time-varying structure to the expelled air associated with conversations.  Phonetic characteristics of plosive sounds like `P' lead to significantly enhanced directed transport, including jet-like flows that entrain the surrounding air. We will show that the transport distance of exhaled material versus time, in the form of three
distinct scaling laws, represents the typical structure of the flow, including (i) a short ($<$ 0.5 m) distance, with large angular variations, where the complexity of language is evident and responsible for material transport in a fraction of second, (ii) a longer distance, out to approximately  1 m, where directed transport occurs driven by individual vortical puffs corresponding roughly to individual plosive sounds, and (iii) a distance out to about 2 m, or even further, where spoken sentences with plosives, corresponding effectively to a train of puffs, create conical, jet-like flows. The latter dictates the long-time transport in a conversation.
Inevitably, there are other complex features, including phonetic structures and the ambient flow, e.g. ventilation, that hopefully will motivate many future studies.

\begin{figure*}[th]
	\centering
	\includegraphics[width=0.8\linewidth,angle=0]{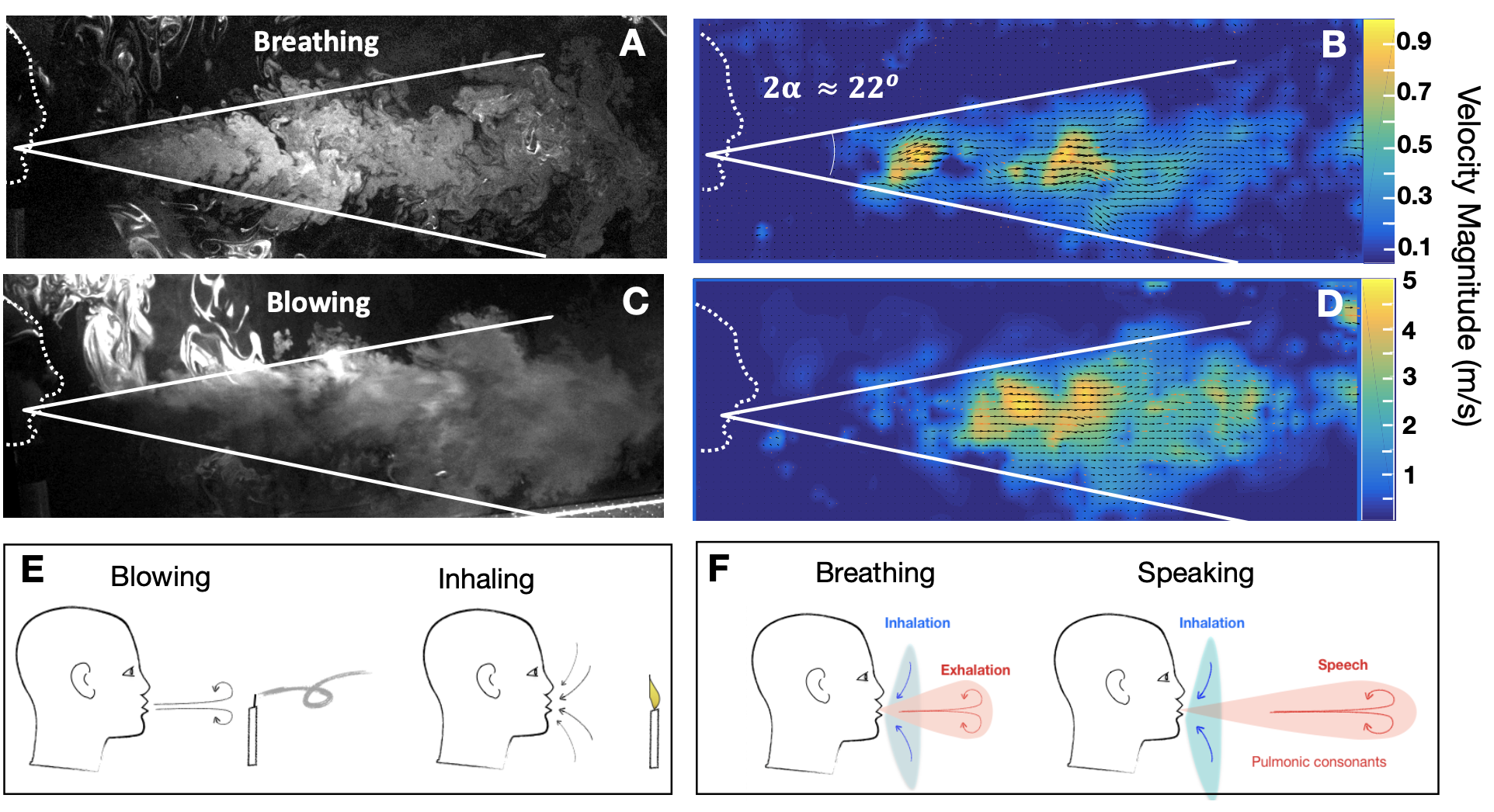}
	\caption{Flow visualization snapshot of exhalation in a laboratory-generated fog and parallel to a laser sheet in two different breathing situations. (A) Calm breathing with (B) the corresponding flow speeds shown with the color code and arrows, and (C) a case of strong blowing with (D) the corresponding velocity field. Notice the much higher velocities associated with blowing. However, the flows in the two cases are qualitatively similar over a sufficiently long period of time of a few seconds and exhibit jet-like features. The field-of-view in all of the images is 1 meter. (E) Sketch of blowing out a candle (or not). (F) Sketch of the qualitative contrast between exhalation and inhalation for breathing and speaking.}
	\label{fig1:CalmBreathing}
\end{figure*}

\section*{Flow Structures of Exhalation and Inhalation: Experiments}

Breathing and speaking are part of our every day activities. We utilize both our mouth and nose. 
We focus on the dynamics of in-flow and out-flow from the mouth since we believe that they are more directed towards a potential facing interlocutor, and we show how some of the features change between breathing and speaking, and are influenced by distinct features of speech, with consequences for transport of exhaled material.

\begin{table}
	\begin{center}
		\caption{\label{tab:peak}Peak flow rates or flow velocities in human breathing and speaking reported in the literature. We assume a typical length scale for the orifice or mouth of diameter $2a=2~\hbox{cm}$ for calculating the Reynolds numbers, $\hbox{Re}=2ua/\nu$, where $u$ is the average speed at the mouth or orifice exit and the kinematic viscosity of air $\nu\approx 1.5\times 10^{-5}$ m$^2$/s. }
		\begin{tabular}{ |c|c|c| } 
			\hline
			& breathing & speaking \\
			\hline
			peak flow rates  & 0.7 L/s \cite{Gupta2010} & $0.3-1.6$ L/s \\ 
			or velocities & 0.5 m/s  \cite{Xu2015}  &  \cite{Gupta2010,Chi2015,Subtelny1966,Isshiki1964,Machida1967,Shadle2010}\\ 
			peak Reynolds numbers& $7\times 10^{2}-3\times 10^3$ & $1\times 10^3-7\times10^3$ \\ 
			\hline
		\end{tabular}
	\end{center}
\end{table}

\subsection*{Orders of Magnitude} 

The typical human adult has a head with approximate radius 7 cm. We may define the characteristic length scale of the mouth, whose shape is approximately elliptical, with the radius $a$ of a circle having the same surface area. Measurements show that the average mouth opening areas are approximately 1.2 cm$^2$ for breathing and 1.8 cm$^2$ (with peak values of the order of 5.0 cm$^2$) for speaking~\cite{Gupta2010}. For an order-of-magnitude estimate of the Reynolds numbers, $a=1~\hbox{cm}$ is chosen. It is perhaps surprising to many that typical air flow speeds are $u\approx  0.5-2~\hbox{m/s}$ (volumetric flow rates $\approx 0.2-0.7 ~\hbox{L/s}$) when breathing and $u\approx  1-5~\hbox{m/s}$ (volumetric flow rates $\approx 0.3-1.6~\hbox{L/s}$) when speaking; see Table~\ref{tab:peak}. When breathing, exhalation and inhalation occur approximately evenly over a cycle with period about $ 3-5$ seconds \cite{conrad1979speech,Gupta2010}, while during speaking the exhalation period is generally lengthened so that 2/3rds or even greater than 4/5ths of the time may be spent in exhalation. 

The local fluid mechanics of exhaled and inhaled flows of speed $u$ are characterized by Reynolds numbers $\hbox{Re}=2ua/\nu$ (the kinematic viscosity of air $\nu\approx 1.5\times 10^{-5}$ m$^2$/s), which have typical magnitudes Re$=O\left (7\times 10^2-3\times 10^3\right )$ when breathing and Re$=O\left (1\times 10^3-7\times 10^3\right )$  when speaking. Inertial effects are expected to dominate these flows, which will also generally be time dependent and turbulent, as discussed below. 

\subsection*{Breathing and Blowing as Jet-like Flows}

We characterize first the nature of breathing and blowing flows (Fig.~\ref{fig1:CalmBreathing}). We set up a laboratory experiment with a laser sheet (1 m $\times$ 2 m $\times$ 3 mm), where no light hits the speaking subject, who sits adjacent to the sheet. A fog machine generates a mist of microscopic aqueous droplets whose large-scale motions are observed with a high-speed camera oriented perpendicular to the sheet. We obtain the velocity field of exhalation (both during breathing and speaking) by observing how the air stream drags and deforms the cloud in the sheet of light using correlation image velocimetry (see typical images in Fig.~\ref{fig1:CalmBreathing}A and C, with details in Materials and Methods). 

The flows are qualitatively similar during breathing or strong blowing (Fig.~\ref{fig1:CalmBreathing}A and C), though the velocity magnitudes can be quite different (Fig.~\ref{fig1:CalmBreathing}B and D). For instance, typical velocities observed in the air flow while breathing with a slightly open mouth ($\sim 1~\hbox{cm}\times 2~\hbox{cm}$) remain of the order of 0.3 m/s to 1 m/s as visible in Fig.~\ref{fig1:CalmBreathing}B (see Movie S1 in Supplementary Information (SI)), while velocities can be as high as a few meters per second in the blowing stream (Fig.~\ref{fig1:CalmBreathing}D) (see Movie S2 in SI). Most significantly, a jet-like, conical structure is visible for the two different situations as depicted by the white lines in Fig.~\ref{fig1:CalmBreathing}A and C, with a cone angle $2\alpha\approx 20^\circ$. We can expect stronger propagation when breathing after exercising, as the volumetric flow rates are increased, which could make breathing in such a case closer to blowing. 
These observations call for comparison for the more complex situation relevant for pathogen transport, which is the case of speaking, where aerosols are produced during speech \cite{AsadiRistenpart2019,AsadiRistenpart2020}. Next, though, we comment on a fundamental asymmetry of exhalation and inhalation.

\subsection*{Asymmetry of Exhalation and Inhalation}

At these Reynolds numbers, we expect exhalation and inhalation to be asymmetric. A reader may be aware that one extinguishes a candle by blowing, but it is not possible to do so by inhalation (Fig.~\ref{fig1:CalmBreathing}E), which is a characteristic of the flows for breathing and speaking.  Long exhalation should produce starting jet-like flows propagating away from the individual over a significant distance of the order of a meter (e.g. Fig.~\ref{fig1:CalmBreathing}A-D), while inhalation is more uniform and draws the air inward from all around the mouth (Fig.~\ref{fig1:CalmBreathing}F); it is this asymmetry that explains the phenomenon related to extinguishing a candle (Fig.~\ref{fig1:CalmBreathing}E). These out-flows are in fact responsible for transporting large droplets and aerosols away from the speaker.

For such inertially-dominated flows, a continuous or long out-flow should be similar to an ordinary jet \cite{list1982turbulent}, and during the initial instants over a time $T$ the propagation distance, while smaller than the naive estimate $L=uT=O(1)~\hbox{m}$ (see below), is still larger than the typical size of the head (e.g. Fig.~\ref{fig1:CalmBreathing}). Moreover, since $L\gg a$,  it follows that, in ordinary circumstances, one breaths in little of what is breathed out.
Wearing a mask (as recommended as a mitigation strategy for COVID-19) should be expected to produce more symmetric flow patterns during exhalation and inhalation, localizing air flow around the face. 

\begin{figure}[th]
	\centering
	\includegraphics[width=0.4\linewidth,angle=0]{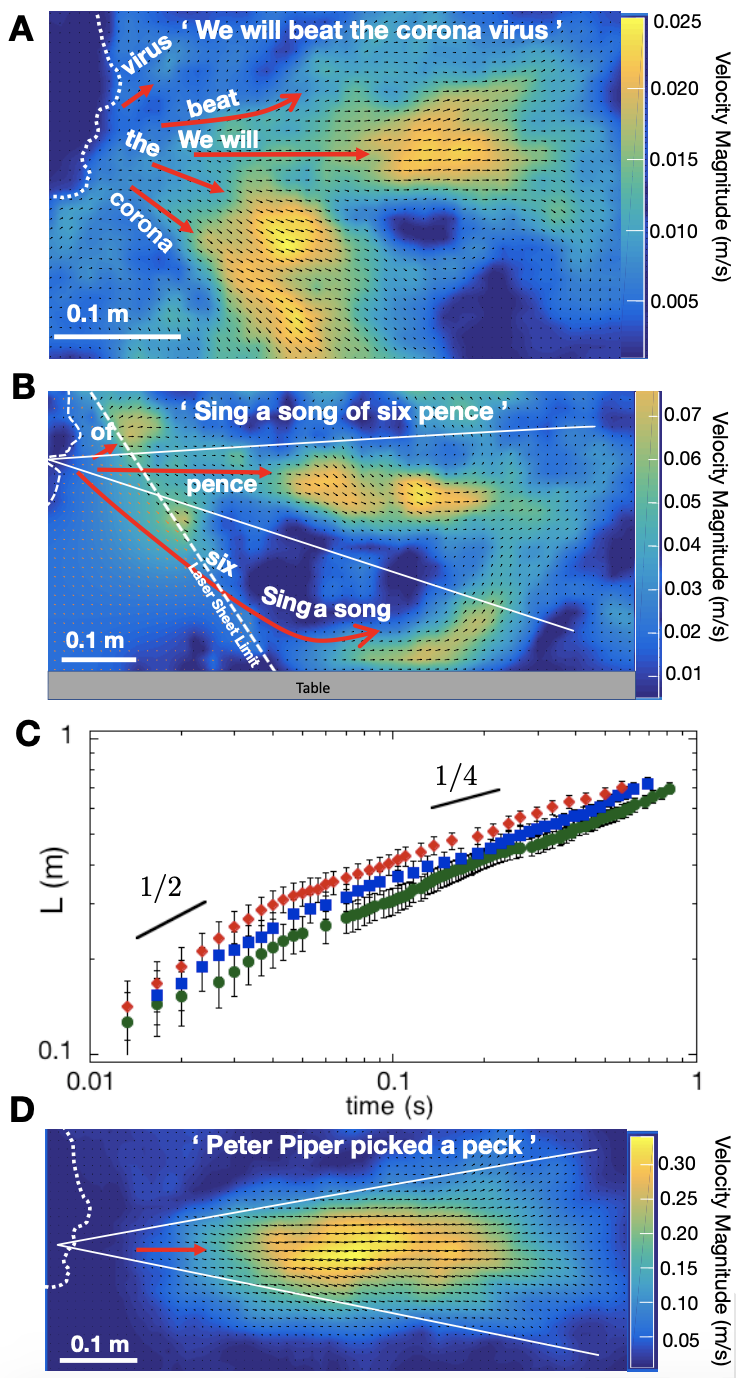}
	\caption{Mean velocity field produced when speaking three different sentences.  A color code illustrates the average speeds but note that single images of the magnitude of speeds are not representative of the true instantaneous velocities, which were estimated from the movies in the SI. (A) `We will beat the corona virus', which is a mixture of vowels, fricatives and plosives. (B) `Sing a song of six pence' (SSSP)~\cite{Subtelny1966}, mainly composed of the fricative `S' except the last word that starts with `P'. (C) The distance travelled by the extremity of the air puff as a function of time when saying `pence' at the end of SSSP for three different runs. (D) `Peter Piper picked a peck' (PPPP)~\cite{Subtelny1966}, which is mainly composed of many plosives `P'.}
	\label{Different sentences}
\end{figure}

\subsection*{Speaking, Plosive Sounds and Jet-like Flows}

Flows exiting from an orifice are well-known to produce vortices, even in the absence of coughing, and these drive the transport about the head, as evident in Fig.~\ref{fig1:CalmBreathing}.
Speaking introduces two further differences: (i) the typical time of inhalation is about 
1/4--1/2 of the exhalation time \cite{conrad1979speech} 
and (ii) language includes rapid pressure and flow rate variations associated with sound productions (plosives, fricatives, etc.), as previously characterized acoustically by linguists \cite{Isshiki1964}. We also note that the stop consonants, or what are referred by linguists as plosives consonants, such as (`P', `B', `K', ... ), have been demonstrated recently to produce more droplets \cite{AsadiRistenpart2020}. In these cases, the vocal tract is blocked temporarily either with the lips (`P', `B') or with the tongue tip (`T', `D') or body (`K',`G'), so that the pressure  builds up slightly and then is released rapidly, producing the characteristic burst of air of these sounds; in contrast, fricatives are produced by partial occlusion impeding but not blocking air flow from the vocal tract~\cite{ChenSpeech}. 

We now visualize flow during speaking, which seems different than breathing as, for instance, when saying a sentence like `We will beat the corona virus', as shown in Fig.~\ref{Different sentences}A (and visible in the Movie S3 of SI). A color code illustrates the average speeds (averaged over the time to say the phrase), but note that these are not representative of the true instantaneous velocities, which in the remainder of this section were estimated from the movies in the SI. Over the approximately 2.5 s to say the sentence, the air flow is more jerky and changes direction depending on the sound emitted.  In this particular case, the sentence contains starting vowels (in `We' and `will') and pulmonic consonants as fricatives (as `V' and `S' in `virus') and plosives (like `B' and `K' in `beat' and `corona'). Three different directions are revealed when averaging the velocity field over the time to say the sentence in Fig.~\ref{Different sentences}A:  `We will beat' being slightly up and to the front with a typical velocity of about 5-8 cm/s, `the corona' being directed downward between $-40^{\circ}$ and $-50^{\circ}$ with higher velocities of almost 8-12 cm/s while saying the two syllables `coro'. Finally, the short air puff associated to `virus' is directed upward at about $50^{\circ}$ with speeds of 5-7 cm/s. We believe that an interlocutor and potential receiver of the exhaled material will be most exposed after a few seconds by the horizontally directed part of the flow, whose velocity reaches, in this case, the ambient circulation speed at about half a meter at most. 

Next, we illustrate a sentence of the same time lapse of about 2.5 s containing many times the same starting fricative `S' as in `Sing a song of six pence'~\cite{Subtelny1966}  with only one starting bilabial plosive sound `P' in the last word: most of the air puffs produced are emitted downward at an estimated angle of $-50^{\circ}$ from the horizontal (and become visible in this sequence only when the air flow hits a nearby table and crosses the laser sheet, see Fig.~\ref{Different sentences}B and Movie S4 in SI). However, a distinct, directed air puff appears in front of the speaker when `pence' is pronounced (Fig.~\ref{Different sentences}B), which propagates forward at initially high speeds of about 1.4 m/s as visible in Movie S4, but decelerates rapidly to $\approx 1 $ m/s at half a meter distance from the mouth; the puff has a speed of 30 cm/s at about 0.8 m (see Movie S4).

These images of typical speech raise the question of the dynamics of individual puffs. In Fig.~\ref{Different sentences}C we report the distance $L$ travelled by the air puff as a function of time $t$ when pronouncing `pence'. The data demonstrates that the starting plosive sounds like `P' induce a starting jet flow, which grows initially for very short timescales of under 10-100 ms as $t^{1/2}$, but rapidly transitions to a slower movement characterized by a $t^{1/4}$ response, typical of puffs \cite{Sangras} and vortex rings \cite{maxworthy1974turbulent}. In fact, when looking at the flow, a vortex ring stabilizes the transport over a distance of almost a meter. This transition between two different dynamics, ending with the dynamics of an isolated puff, is also measured in coughs~\cite{BourouibaBush2014}.

In contrast, when we speak a sentence with many `P’ sounds, such as `Peter Piper picked a peck' (PPPP)~\cite{Subtelny1966}, as illustrated in Fig.~\ref{Different sentences}D, the distribution of the average velocity field approaches that of a conical jet with average velocities of tens of cm/s and over long distances of about a meter. Peak velocities are seen at the emission of the sound `P' with values close to 1.2-1.5 m/s (Movie S5 in SI). This more directed flow situation shares features of breathing and blowing and thus material will be transported faster and further than individual puffs. But, unlike breathing, we believe that this distinct feature of language is more likely to be important for virus transmission since droplet production has been linked to the types of sounds \cite{AsadiRistenpart2020}. 

It should be evident that language is complicated (Fig.~\ref{Different sentences}A, B).  Given the possibility of asymptomatic transmission of virus by aerosols during speech, we have focused on the phrases in language, those usually containing plosives, that produce directed transport in the form of approximately conical turbulent jets (Fig.~\ref{Different sentences}D, and also see Figs.~\ref{fig3:BreathingVsSpeaking} and \ref{fig5:SheetExperiments} below). 

In addition, to see that thermal effects are small until the jet speeds are reduced to closer to ambient speeds, consider the Richardson number $Ri = \frac{g\frac{d\rho}{dz} }{\rho \left (du/dz\right )^2}$. So approximately $Ri\approx \frac{\Delta \rho}{\rho} \frac{g \Delta z}{\left (\Delta u\right )^2}$. For a $15^\circ\hbox{C}$ degree temperature change in air, $\frac{\Delta \rho}{\rho} \approx 0.05$, so with 
$\Delta u\approx 0.5$~ m/s and a length scale say $\Delta z\approx 0.1$~ m (which is relatively large), we find $Ri\approx 0.2<1$. The thermal effects should be expected to be important at longer distances where the jet speed is reduced (usually where the ventilation may also matter) or if a mask is used which decreases the flow speed substantially. 

\begin{figure*}[tbp]
	\centering
	\includegraphics[width=0.9\linewidth,angle=0]{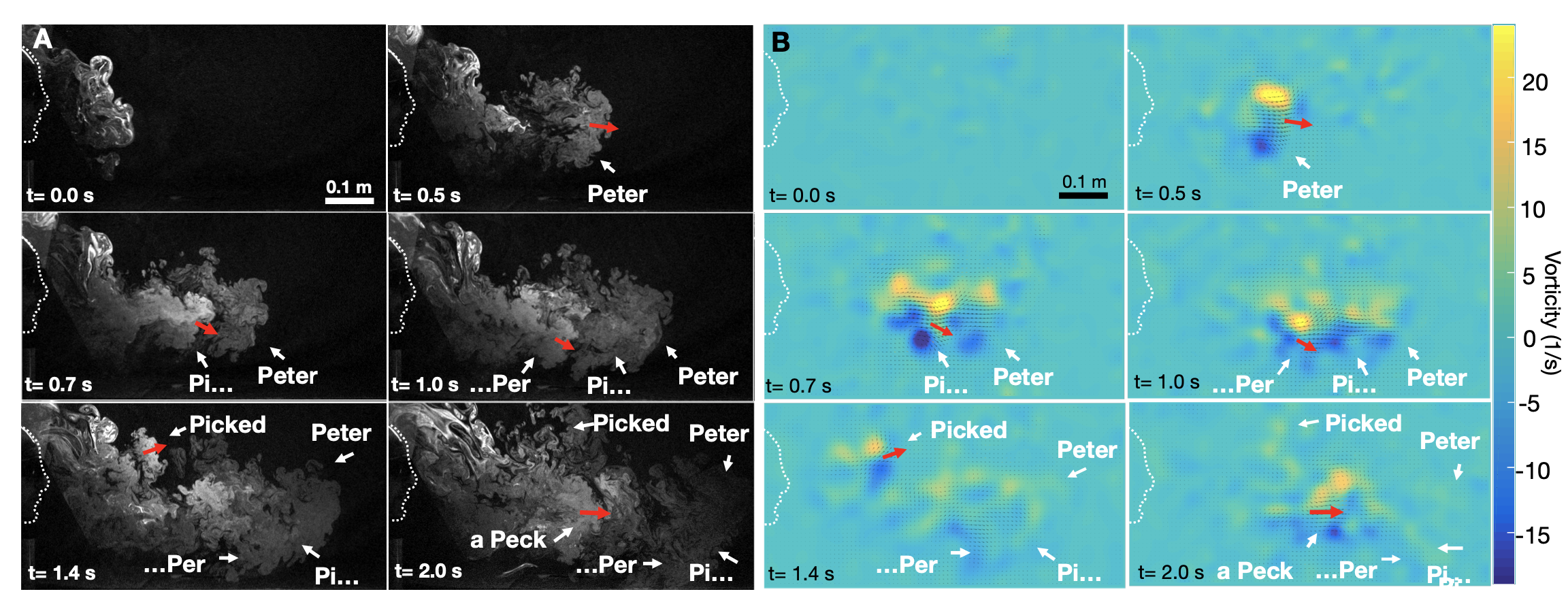}
	\caption{Time evolution of the flow from a phrase, `Peter Piper picked a peck' (PPPP), with many plosives, spoken parallel to a laser sheet. The speaker is indicated by the dotted curve to the left. to the (A) Flow visualization with the individual plosives identified. (B) Vorticity field with individual vortices clearly visible for each plosive `P' pronounced in the sentence. Notice the interactions between the first vortices, as well as the different upward angle of the vortex produced when the syllable `Pi' is pronounced in `picked'.}
	\label{Plosives of PPPP}
\end{figure*}

We document the distinct role of the individual plosives in the phrase `Peter Piper picked a peck' (PPPP) with the time-lapse images displayed in Fig.~\ref{Plosives of PPPP}A (see also Movie S5 in SI). By  performing correlation image velocimetry to calculate the vorticity field $\mathbf{ \omega}=\nabla\wedge{\bf u}$, where $\bf u$ is the in-plane velocity field, as shown in Fig.~\ref{Plosives of PPPP}B, we could follow the vortical structures created by the pronunciation of `P's in PPPP. Vortices shedding from the mouth are clearly visible,  interact, and survive downstream where they easily reach the meter scale. The transition from puff-like dynamics associated to single plosives and the development of turbulent jet-like flow during longer sentences seems to be associated with the sequential accumulation of `puff-packets' pushing air exhaled from the mouth. We will explore this transition in more detail using the  numerical simulations below.

\section*{Modeling}

To assist with the interpretation of the experimental results just presented, and the numerical results we will report below, for completeness we summarize a few results of well-known mathematical models. 

\subsection*{Characteristic Features of a Steady Turbulent Jet}

In a high-Reynolds-number steady turbulent jet, it is of interest to characterize the volume flux, linear momentum transport and kinetic energy transported by the jet, as well as the entrainment of the surrounding air that dilutes the jet~\cite{LeeChu}. These properties also help to understand the fluid dynamics of breathing and speaking. There are at least three significant conclusions that characterize the flow: (i) Denoting the direction of the jet as $x$, the typical axial speed of the jet as $v(x)$, and its cross-sectional area as $A(x)$, in a steady jet issuing into an environment at a constant pressure, the flux of linear momentum is constant, or $v^2A=\hbox{constant}$. If the exit flow near the mouth is characterized by a speed $v_0$, volumetric flow rate $Q_0$ and area $A_0$, we conclude that $v(x)/v_0 = \left (A_0/A(x)\right )^{1/2}<1$. For a conical jet-like configuration of angle $\alpha$ (Fig.~\ref{fig1:CalmBreathing}), then beyond the mouth $A(x) \propto \left (\alpha x\right )^2$.  (ii) The corresponding  volume flux $Q=vA $, so that the out-flow leads to a volume flux $Q/Q_0=\left (A(x)/A_0\right )^{1/2}>1$, i.e., there is entrainment of the surrounding air into the jet, which is an important feature of mixing of the surroundings. (iii) Any material expelled from the mouth with concentration $c_0$ is reduced in concentration as the jet evolves, with $c(x)/c_0 = Q_0/Q(x)$. 
Since the jets are approximately conical, then the above results predict that the characteristic quantities vary with distance as $v(x) \propto \left (\alpha x\right )^{-1}$, $A(x) \propto \left (\alpha x\right )^2$, $Q \propto \alpha x$ and $c \propto \left (\alpha x\right )^{-1}$. Although these arguments are based on the assumption of a steady jet, we shall now see that they apply approximately to the unsteady features of speaking on the time scale of many cycles and far enough from the mouth or exit of an orifice.

\subsection*{Starting Jets and Puffs} A jet formed by the sudden injection of momentum out of an orifice is referred to as a starting jet. Such flows reach a self-preserving behavior some distance downstream of the source, where the penetration distance grows over time like $L\propto t^{1/2}$~\cite{Sangras,Derrick2009}; see also equation (\ref{eq:L_vs_t}) below.

On the other hand, a rapid release of air, or puff, injects a finite linear momentum into the fluid, e.g. Fig.~\ref{Different sentences}. For the inertially dominated flows of interest here, the linear momentum of the puff is conserved, so that the distance travelled is
$L\propto t^{1/4}$~\cite{Sangras,Derrick2009}, similar to interrupted jets, i.e., starting jets when the flow is suddenly stopped.

However, during breathing or speaking, the interrupted jet and the puffs are released one after the other and interact with each other in front of the source, as illustrated by Fig.~\ref{Plosives of PPPP}. The jet is neither continuous like in starting jets nor isolated like in classical puffs. What is then the dynamics of such a ``train of puffs''? In the next section, we use numerical simulations to investigate the dynamics of puff trains and quantify their growth in space and time.

\section*{Three-dimensional Numerical Simulations: Characterizing the ``Puff Trains'' of Breathing and Speaking}

To explore quantitatively the various flows we have introduced above, we report 3-D simulations of the incompressible Navier-Stokes equations (the flow speeds are much smaller than the speed of sound). 
To highlight the dynamics of breathing and speaking, simulations are driven by representative time-periodic flow rate variations \cite{Subtelny1966} from an elliptical orifice comparable to a large open mouth (of radii 1 cm $\times$ 1.5 cm). Speaking produces relatively high-frequency changes to the volume flow rate (or fluid speed) during exhalation, though the variations are much smaller than sound frequencies; we do not study the initial formation of the sounds of speech at the glottis~\cite{MittalErathPlesniak}. Furthermore, as we have seen above, natural plosive sounds also create special characteristic features that we investigate. Nevertheless, it has to be stressed that the simulations are a model and lack the phonetic complexity introduced by the tongue and the cavity of the mouth, yielding flows directed in front of the mouth only. 

\begin{figure*}[tbph]
	\centering
 \includegraphics[width=0.9\linewidth,angle=0]{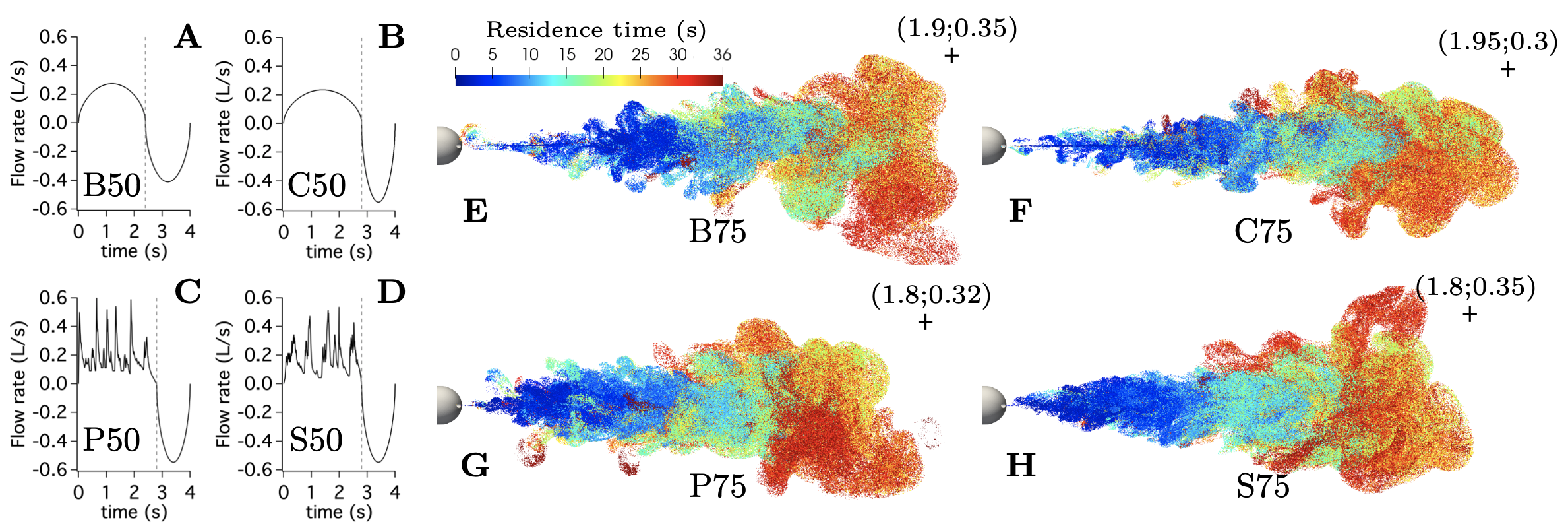}
	\caption{Numerical simulations of periodic breathing versus speaking signals for cycles of 4.0 s. The jets issue from a sphere with an open elliptic orifice of semi-axes 1.0 cm and  1.5 cm. (A-D) volumetric flow rate signals for cases with 0.5~L per breath (hence the `50' in the labels), where the vertical dashed lines mark the separation between exhalation and inhalation. (A) Case B50 is a breathing-like signal with 2.4 s exhalation and 1.6 s inhalation. (C) Case P50 and (D) case S50 are speaking signals sampled from \cite{Subtelny1966}, and recorded during articulation of `Peter Piper picked a peck' and `Sing a song of six pence', respectively, with speaking time of 2.8 s and a 1.2 s inhalation. The P50 and S50 signals have been adjusted to 0.5 L per breath. (B) Case C50 is a calm signal of the same macroscopic characteristics as P50 and S50, but with a smooth signal similar to B50. Three series of simulations have been performed at different volumes per breath, i.e., 0.5 L, 0.75 L and 1.0 L per breath. For the simulation of `Peter Piper picked a peck' for instance, the simulations at 0.75 L and 1.0 L per breath are referred to as P75 and P100, and are obtained by multiplying the input flow rate signal of P50 by 1.5 and 2.0, respectively. (E-H) Examples of jets obtained for cases B75, C75, P75 and S75 after 9 cycles (36 s), as visualized by perfect tracers issued from the mouth and color-coded by their residence time in the computational domain (dark blue tracers were exhaled during the last cycle). The scale is the same for all plots.  For each case, a point marked by a '+' is positioned to indicate the axial and radial extent of the jet. The $x$ and $y$ coordinates of the point are reported as ($x$;$y$) in the figures (E-H). The sphere representing the head is shown to the left in each panel.}
	\label{fig3:BreathingVsSpeaking}
\end{figure*}

\subsection*{Contrasting Four Situations of Exhalation}  

We contrast four situations with comparable period and given volumes exhaled and inhaled, with zero net out-flow over one cycle  (Fig.~\ref{fig3:BreathingVsSpeaking}A-D): (i) normal breathing with a 4-second period split into intervals of exhalation (2.4 s) and inhalation (1.6 s); (ii) a breathing-like signal but with a (slow) speaking-like distribution of exhalation (2.8 s) and inhalation (1.2 s), (iii) a spoken phrase, `Sing a song of six pence'~\cite{Subtelny1966}, and (iv) a phrase with many plosive sounds, `Peter Piper picked a peck'~\cite{Subtelny1966}. We either ran 1-cycle simulations using the flow rate profiles over a single period, followed by no further out-flow, 
to quantify a single ``atom'' of breathing and speaking, and for many periods (or cycles) to understand how the local environment around an individual is established and changes in time. Different volumes of exhalation typical of speaking, from $0.5 - 1$ L per breath, were studied (see the full table of runs in the SI, Table~S1).

The results of simulations of these different flow rate profiles are shown in Fig.~\ref{fig3:BreathingVsSpeaking}E-H for an exhaled volume  of $0.75$~L per breath. To visualize the flow, tracers injected at the in-flow are shown, color-coded by the residence time of the tracers. For every case, a conical jet flow is produced, with similar cone angles as well, which is reminiscent of typical features of turbulent jets studied in laboratory experiments and many applications, e.g. \cite{LeeChu,Choutapalli2009}; see also Figs.~\ref{fig1:CalmBreathing}-\ref{Different sentences}. Qualitatively, we observe that breathing produces a jet with an axial flow comparable to speaking, which some may find surprising. Jet lengths in particular are very similar, despite a factor of 2.6 in the peak flow rate of cases P75 and C75 for instance (see Table~S1). 
The phrase with plosives produces qualitatively a rougher jet (Fig.~\ref{fig3:BreathingVsSpeaking}G) due to the ejection of vortex rings away from the main jet and vortex interactions. Speaking jets (P75 and S75) yield the largest cone angles and consequently an axial extent somewhat reduced compared to breathing (B75 and C75). Short high-speed puffs associated with speaking thus seem to increase the jet entrainment, but do not enhance the long-range transport in the axial direction. 

For all cases, even those with complex phonetic characteristics, we observe that the resulting jets display many of the features of a turbulent jet, which leads to transverse spreading and mixing of the exhaled contents with the environment. These features actually build up over the continual cycles of exhalation and inhalation in both breathing and speaking. Particle residence time (Fig.~\ref{fig3:BreathingVsSpeaking}E-H) notably show the progressive formation of the jet. However, a striking feature is the absence of obvious signature of the flow pulsation in the far field. From the global point of view, all computed jets, whatever the details of the in-flow signal, are similar to steady turbulent jets away from the immediate vicinity of the mouth.

\begin{figure*}[h]
	\centering

  \includegraphics[width=0.8\linewidth,angle=0]{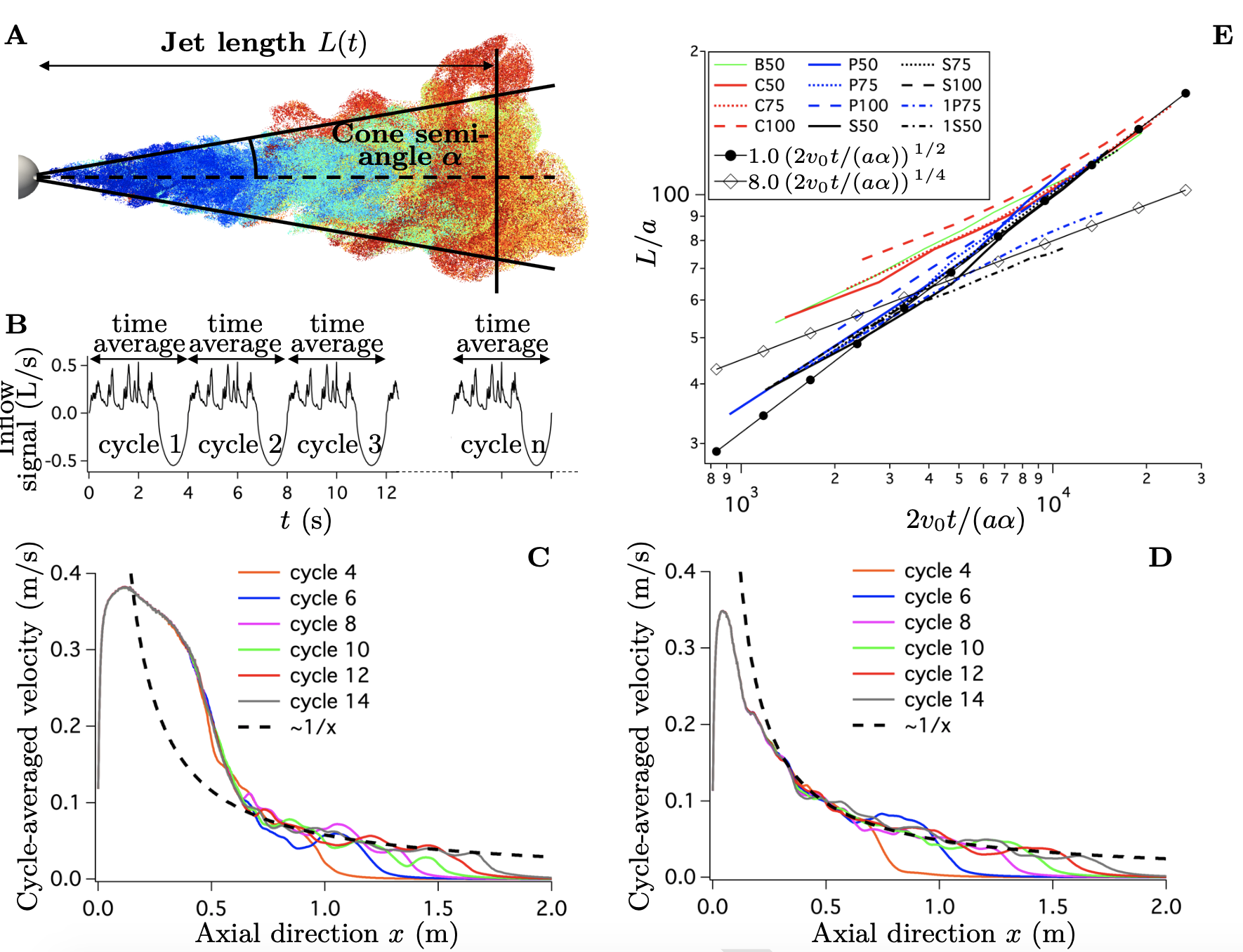};

	\caption{Jet characteristics in the simulation of the breathing and speaking signals shown in Fig.~\ref{fig3:BreathingVsSpeaking}A-D. (A) Example of calculation of jet length $L$ and angle $\alpha$ for S75, based on the emitted tracer particles color-coded here by residence time, as described in Fig.~\ref{fig3:BreathingVsSpeaking}E-H. $L(t)$ is such that 90\% of the particles are located upstream of $x=L$ at time $t$. The cone angle $\alpha$ is calculated to enclose 90\% of the particles is a cone passing through the mouth exit (radius is 1 cm at $x=0$). This angle is verified to remain stable with time after the initial cycles.  (B) Principle of the calculation of the cycle-averaged velocity fields presented in (C-D): The velocity is time-averaged independently over each cycle. (C-D) Progressive formation, along increasing cycles of exhalation and inhalation, of a turbulent jet-like velocity profile $v(x)$ in the far field. Examples of cases C50 (C) and S50 (D):  cycle-averaged axial velocity along the $x$ axis from the mouth exit to $2.0$ m downstream, for different cycles, extending to 14 cycles or 56 s. The black dashed line is the $v(x)\propto 1/x$ scaling, plotted here as a guide for the eye, which is suggested by a model of a steady turbulent jet.
		(E) Evolution of the non-dimensional jet length $L/a$ as a function of $2 v_0 t/(a\alpha)$ (see Eq.~\ref{eq:L_vs_t}), with $a\approx 1.22$ cm the equivalent radius of the mouth exit and $v_0$ the average axial speed during exhalation for the different simulations. Two power laws are plotted as a guide for the eyes to assess the evolution of $L$ with time. Raw data for $L(t)$ is plotted in Fig.~S2.}
	\label{fig4:NumericalData}
\end{figure*}

\subsection*{Quantifying the Jets}

We ran simulations for three different flow rate signals and exhaled volumes (0.5, 0.75 and 1 L/breath) to understand the characteristics of the flows generated by multi-cycle breathing and speaking. Because the flows are time-varying and turbulent we quantified the cone half angle $\alpha$ by determining the angle inside of which reside 90\% of the exhaled tracer particles (Fig.~\ref{fig4:NumericalData}A).  The included angles differed from case to case, but were of the order of $10-14^\circ$ (see Table~S1). The typical jet lengths, $L(t)$, were also calculated based on the criterion that 90\% of the tracers are located upstream of $x=L$ at time $t$. 
Raw data of $L(t)$ are presented in the SI, Fig.~S2, and the jet angles are reported in Table S1.  
First, higher mean flow rates (exit speeds) produce longer lengths, as expected. 
For a given exhaled volume per cycle, different types of exhalation produce comparable jet lengths, as suggested by the qualitative analysis of Fig.~\ref{fig3:BreathingVsSpeaking}E-H. Modulation of the in-flow signal (cases P and S)  systematically tends to increase the lateral growth of the jet, increasing the jet angle and decreasing the jet length. 

\subsection*{A Train of Puffs} 

We ran the multi-cycle simulations over many periods to quantify the development of the transient velocity field.  In order to filter the turbulent fluctuations that prevent direct comparisons of the velocity fields as a function of time, we
performed time averages over each period (see Fig.~\ref{fig4:NumericalData}B) to produce an approximate profile for the distribution of axial speeds in the exhaled jet.
In the far field, though time varying, breathing and speaking may be viewed as periodic processes where the time scales are much longer than an individual period. 
Moreover, we have already explained that inhalation has little effect on exhalation because of the differences expected of high-Reynolds-number motions. Indeed, when we plot the axial speed as a function of axial distance we find that for each period of exhalation, the axial velocity falls along the curve $v(x) \propto x^{-1}$ for both speech and breathing, shown, respectively, in Fig.~\ref{fig4:NumericalData}C and D. Not only does the head of the jet evolve as that of a starting jet, but the whole flow downstream of a certain distance from the mouth behaves similarly to as a steady  turbulent starting jet. This is particularly striking as the near-mouth flow is laminar and completely different from a steady jet (Fig.~\ref{fig4:NumericalData}C-D). Thus, at the Reynolds numbers characteristic of breathing and speaking, a train of puffs transitions to a turbulent, jet-like flow that dominates the transport associated with breathing and speaking. 

\subsection*{A Diffusive-like, Directed Cloud of Exhaled Air}

For growing jets at constant angle, we can estimate the spreading of the cloud with time. The time $t$ it takes to reach an axial distance from the orifice, or the mouth, is estimated by 
\begin{equation}
t=\int_0^{L(t)} ~\frac{{\rm d}x}{v(x)}\approx \frac{\alpha L^2}{2v_0a}
\label{DirectedJe1}
\end{equation}
or (using $a$ to make the equation non-dimensional)
\begin{equation}
\frac{L(t)}{a} \approx \left (\frac{2 \, v_0 \, t}{a\, \alpha}\right )^{1/2}.
\label{eq:L_vs_t}
\end{equation}
The scaling from this equation is that expected for starting jets~\cite{Sangras}.

The theoretical prediction for the length of the exhaled air column for a starting jet (Eq.~\ref{eq:L_vs_t}) is then compared to the non-dimensional numerical data in Fig.~\ref{fig4:NumericalData}E. The scaling captures quantitatively the trends provided that $v_0$ is defined as the average speed at the orifice exit (mouth) during exhalation. The peak velocity is not relevant: strikingly, the details of the flow rate signal do not impact the scaling, but only influence the spreading angle of the jet. In addition, for 1-cycle simulations (1P75 and 1S50), we recover that the whole exhaled material acts as a unique large puff~\cite{Sangras}, and $L\propto t^{1/4}$ is obtained, which is consistent with the experiments (Fig.~\ref{Different sentences}C).

These results allow the quantification of concentration of exhaled material in the far field. From the previous results, we expect the concentration field of the exhaled cloud is quasi-steady and falls off with distance, $c(x)/c_0\propto a/\left (\alpha x\right )$. Note that for $a =1~\hbox{cm}$, $\alpha = 10^\circ$, and $L=2 ~\hbox{m}$ (the six-foot rule), then for directed jets the concentration of any exhaled material has fallen off by a factor of $\left( a/\left (\alpha L\right) \right)\approx 0.03$.  Typical dilution levels of 0.04-0.05 have been found in the different  simulations at $1.5$ m, which is consistent with this estimate. It is evident that this result is not an especially large dilution and the concentration is much larger than might be estimated based on a model of diffusion from a sphere.

\begin{figure}[th]
	\centering
	\includegraphics[width=0.6\linewidth,angle=0]{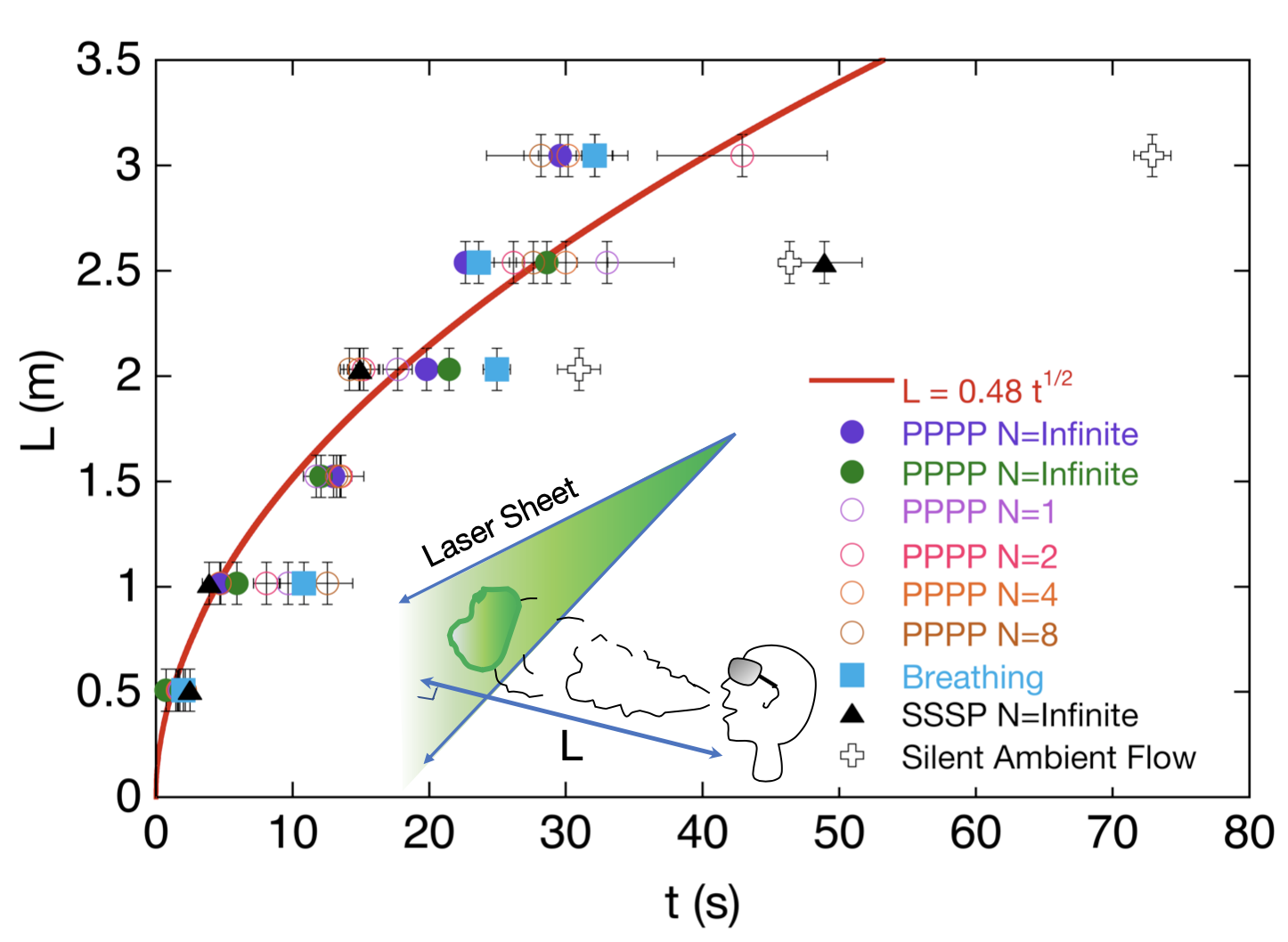}
	\caption{Speech propagation distance $L$ to a laser sheet versus  time $t$ while saying `Peter Piper picked a peck' (PPPP) and 'Sing a song of six pence' (SSSP); $N$ indicates the number of times the sentence has been repeated (with 1 second inhalation in between) before total silence and $N=\hbox{Infinite}$ means the sentence has been pronounced until the sheet was reached. The solid line represents a fit with a $t^{1/2}$ power law. `Silent ambient flow' (crosses) refers to the control case where the fog is convected by the ambient flow alone.}
	\label{fig5:SheetExperiments}
\end{figure}

\section*{Experimental Characterization of the Spreading}

To complement the numerical simulations and to further characterize the propagation of the exhaled jets we placed a laser sheet perpendicular to a speaker (Fig.~\ref{fig5:SheetExperiments} inset). We measured the time $t$ for the laser sheet to be visibly disturbed when placed a distance $L$ in front of the speaker, who said the sentence `Peter Piper picked a peck' (PPPP) $N$ times. The data of $L(t)$ (circles), including breathing (diamonds), along with the background flow (crosses) and SSSP (triangles), is shown in Fig.~\ref{fig5:SheetExperiments}.
For the plosive phrase, for all $N$, we observe good agreement with the prediction $L\propto t^{1/2}$ (the solid curve) obtained by representing the far field of the out-flow from speech as a steady turbulent jet. We note that the data for
SSSP at long times deviates from the theory, perhaps because of  intermittency introduced by only an occasional plosive. 
The prefactor of the fit of 0.48 obtained for the PPPP data together with breathing data compare well to the scaling law given in Eq. \ref{eq:L_vs_t}: considering a mouth on average opened at $a=1-2$ cm while saying PPPP, a typical air speed $v_0\sim 1.2-1.5$ m/s at the exit of the mouth when saying the plosives `P' (see Movie S5) and an angle of $\alpha=10^\circ$, we obtain $\left (2v_0 a / \alpha\right )^{1/2} \approx 0.37-0.59$.

Note that there is a weak ambient flow in the laboratory throughout the experiments. The speed of the ambient flow is $O(0.05 ~\hbox{m/s})$, where the initial speeds of the experiments with many plosives and words are $O(0.5 ~\hbox{m/s})$. Therefore, the ambient flow does not significantly affect the initial spreading of the puff (e.g. $L < 1$~ m), but does affect the transport of the puff during spreading and deceleration (e.g. $L > 2 ~\hbox{m}$). The ambient flow introduces uncertainty to these experiments of about 20 \% (Fig.~\ref{fig5:SheetExperiments}). On the other hand, the existence of an ambient flow is ubiquitous and our results can provide a means to estimate the cross-over between speech-dominated transport and ventilation-dominated transport. Though we do not pursue the topic here, the effect of the ambient flow is an interesting and important problem for further investigation.

\section*{Discussion}

We believe that this work is one of the first to quantify the fluid dynamics of the environment about the head of a person while breathing or speaking. Some features are relatively easy to understand, such as the natural asymmetry of exhalation and inhalation, which contribute to the ``cloud'' of exhaled air being continually pushed away as it mixes with the environment. 
Taken together, our results have identified three typical regions of transport associated with conversations (i.e., a series of sentences) that contain plosives: (i) Less than about 50 cm from the speaker, exhaled material is delivered in a fraction of a second with flows directed upwards (about  40$^\circ$ from the horizontal), downwards  about  40$^\circ$ from the horizontal) and directly in front (especially the bilabial plosives), where the latter regime obeys a $t^{1/2}$ starting-jet power law; (ii) out to about 1 m, longer, though slower, transport occurs driven by individual vortical puffs created by syllables with single plosives, where the time variation follows a $t^{1/4}$ power law; (iii) finally, out to about 2 m, or even further, due to an accumulation of puffs, the exhaled material decelerates to about a few cm/sec and becomes susceptible to the ambient circulation (in our ventilated lab). In this last regime, we discovered that the series of puffs, from plosives in a spoken sentence, produces a conical, jet-like flow, again similar to a starting jet, with a $t^{1/2}$ power law.

In the absence of significant ventilation currents, or air motions driven by other speakers, we have seen that often the exhaled cloud will largely be in front of the speaker, with a modest angle as shown in this paper. 
The dynamics of ``puffs'' associated with individual breaths or sounds have a distinct dynamics with the very early-time formation phase having a distance that scales with $t^{1/2}$ after which the puff advances a downstream distance that varies with $t^{1/4}$; these dynamics are common to starting jets of all types (e.g.~\cite{Sangras}), including coughs~\cite{BourouibaBush2014}.
However, speech is similar to a train of puffs, effectively generating a continuous turbulent jet, which mimics many of the features of exhalation in breathing and speaking, where the local exhaled cloud increases in size approximately as $L\propto \left ( v_0 t\right )^{1/2}$; both longer times and larger flow velocity (or increased breathed volume in the case of exercise for instance) increase the affected environmental volume. Moreover, the droplet emission rate (number of droplets per time) increases with louder speech~\cite{AsadiRistenpart2019}.
With social situations in mind, in hindsight, it should perhaps not be surprising that droplet and aerosol generation, and possible virus transmission, are enhanced during rapid and excited speech during parties, singing events, etc. \cite{ChoirPractice,FitnessClass}

The results presented in this paper do not account for some real features, e.g. movement of the head or trunk of the speaker and the influence of background motions of the air due to the ventilation. 
There is obviously much to be done to quantify the many details and nuances, especially as the different sounds in speech  produce vortical structures of different strengths that influence the spread (axial and transverse) of the exhaled jet. 

The authors are not trained in public health, nor have professional standing in the public health arena, so we should be cautious in conclusions to be drawn from our results regarding social distancing guidelines. Nevertheless, there are general results that can be extracted from this work. Our results show that typical airflow speeds at 1 -- 2 m distances from a speaker are typically tens of centimeters per second. This means that the ambient air current may be dominant at such distances from a speaker, which makes the definition of guidelines difficult. When thinking about quantitative features to discuss social distancing guidelines (six feet, approximately 2 m,  in the United States or 1 m in the World Health Organization's interim guidance published on June 5, 2020 \cite{WHO}), both spatial and temporal characteristics matter, e.g., during conversations, the time spent in front of a speaker, and the distance from the speaker, are needed to define an estimate for the dose of virus received; the dose is proportional to the concentration of aerosol at that distance. Based on the experimental and numerical results reported in this paper, exhaled materials reach  0.5 -- 1 m in a second during normal breathing and speaking, and in fractions of a second in the case of plosive consonants (Figs.~\ref{fig1:CalmBreathing}-\ref{Plosives of PPPP}). If one is directly in the path of the speaker, then at 2 m and within about 30 seconds, the exhaled materials are diluted to about 3\% of their initial value. However, more extended discussions, and meetings in confined spaces, mean that the local environment will potentially contain exhaled air over a significantly longer distance. It follows that in conversations longer than 30 seconds it is better, in our opinion but based on the results in this paper, to move beyond  2  m of separation, and to stand to the side of a speaker, e.g., outside of a cone of 40-50 degrees (half angle), further reduces possible inhaled aerosol.   Most significantly, our results illustrate that 2 m, or six feet, does not represent a ``wall’’, but rather that behavior can help minimize risk by increasing separation distances and relative position for longer conversations when masks are not used.

\section*{Concluding Remarks}

We have provided a quantitative framework to describe a fundamental mechanism of transport that can be generalized to many pathogens.  Obviously, much remains to be done for understanding the fluid mechanics associated with simple human activities, i.e., breathing and speaking. Similar ideas apply to other mammals, though the scales are different between a bat, a bird or a cow. Furthermore, many pathogens might have adapted to use the respiratory systems of humans and other mammals as an efficient transport mechanism. Our work will help better understand virus transmission in mammals, which can have catastrophic consequences in nature or affect the food supply. Building on the understanding of the fluid dynamics of viral and pathogen transmission we believe it will be possible to design potential mitigation strategies, in addition to masks, and vague social distancing rules, and link to poorly understood issues of viral dose~\cite{Rabinowitz} to better manage societal interactions prior to introduction of a vaccine. We invite researchers to combine the full aerodynamics of sound production, including the different phonetic characteristics, and even sound generation in animals, with droplet formation from saliva and mucus to better understand and describe how airborne pathogen biology is adapted for this mode of transport and transmission. 

\begin{acknowledgements}
We thank the NSF for support via the RAPID grant CBET 2029370 (Program Manager is Ron Joslin). M.A. thanks the IRN ``Physics of Living Systems'' (CNRS/INSERM) for travel support, as well as K. Meersohn for pointing out the importance of plosives in almost all languages of the world. S.M. thanks V. Moureau and G. Lartigue (CORIA, UMR 6614) and the SUCCESS scientific group for providing YALES2, which served as a basis for the development of YALES2BIO. Simulations with YALES2BIO were performed using HPC resources from GENCI-CINES (Grant No. A006 and A0080307194) and from the platform MESO@LR. S.M. acknowledges the LabEx Numev (convention ANR-10-LABX-0020) for support for the development of YALES2BIO. We thank A. Smits for loaning the fog machine and P. Bourrianne and J. Nunes for help measuring flow rates during breathing.
\end{acknowledgements}

\appendix

\section{Speaker}
Due to difficulties imposed by the pandemic, only one subject could enter the lab and participate in the experiments. The subject volunteered for the study, is male and 44 years old, with no known physical conditions. The study was approved by the Princeton University IRB (protocol \# 12834). The subject provided informed consent.

\section{Flow visualization}
In the laboratory experiments, a point-wise laser light (wavelength $\lambda = 532$ nm , 1 W power, DPSS DMPV-532-1, Del Mar Photonics) passes through a concave cylindrical lens (focal length $F = -3.91$ mm) and spreads to form a laser sheet about 2 m in length and 1 m in height.
The mean thickness of the laser sheet is approximately 3 mm.
To maintain safe use, the laser light shines from above so that no light hits the speaker who sat adjacent to the sheet. Laser safety glasses were worn by the speaker.

The flow is seeded by a fog machine (Mister Kool by American DJ), which uses a water-based juice (Swamp Juice by Froggys Fog) and generates droplets with diameters of about one micrometer.
The fog can last for tens of minutes and no notable sedimentation of the droplet is observed throughout the course of the experiments.
Therefore, the droplets can track the local flow, effectively as passive tracers.
Images are captured via a high-speed camera (v7.3, Phantom) with frame rate $f = 300$ fps (frame per second).
However, we note that there is inevitable background flow in the experiments due to the droplet emission by the fog machine, as well as the natural ventilation in the room.
Specifically, the background flow is of the order of $O(1)$~cm/s and moves from the left to the right in the experiments reported in the main text (e.g., Fig.~\ref{fig1:CalmBreathing}) and only slightly enhances the propagation of the jets.
Although we do not pursue it here, the effect of the background flow due to ventilation on the transport of the out-flows from breathing and speaking is an interesting and fundamental problem for future investigations.

A similar setup is used when speaking a distance $L$ in front of a laser sheet to determine the axial structure of the out-flows, e.g., the measurement presented in Fig.~\ref{fig5:SheetExperiments}.
The laser sheet is perpendicular to the flow and the camera is perpendicular to the laser sheet.

\section{Correlation image velocimetry}
In order to quantify the structure of the jets from breathing and speaking, the seeded image sequences captured on video are processed using PIVlab~\cite{thielicke2014pivlab}.
The cross-correlation method is applied to the image sequences to measure the local velocities in the particle image velocimetry (PIV) analyses.
Square interrogation windows of 16 pixels $\times$ 16 pixels (approximately 2 cm $\times$ 2 cm) with an overlap step of 50 {\%} (8 pixels, 1 cm) are used to obtain the velocities, e.g. those presented in Fig.~\ref{fig1:CalmBreathing}B.

\section{Numerical simulations}
The computations are performed with the in-house flow solver YALES2BIO~\cite{Zmijanovic:2017,Nicoud:2018,Moureau:2011a,Moureau:2011b,Malandain:2013} (\url{https://imag.umontpellier.fr/~yales2bio/}). These are large eddy simulations~\cite{sagaut2006large}, which are well suited to study transport in turbulent flows, in particular in the context of speech production~\cite{Derrick2009}. In addition, they are well adapted to intermittent/transitional regimes~\cite{Zmijanovic:2017,Nicoud:2018}. The spatially filtered, incompressible form of the Navier-Stokes equations are solved. The so-called sigma model~\cite{Nicoud:2011} is used to treat the effect of the numerically unresolved scales on the resolved scales. Particles are injected into the flow to characterize the jets issuing from the orifice (mouth). They are perfect Lagrangian tracers displaced at the local fluid velocity, and do not affect the flow. In the simulations buoyancy effects are not considered; the temperature, density and dynamic viscosity are constant. The geometry of the model of the mouth  remains constant over time and does not depend on the type of in-flow signal (breathing or speaking). The mouth opening is an ellipse of semi-axes 1.0 cm and 1.5 cm, which corresponds to the upper limit of the range of mouth surface area observed during speaking~\cite{Gupta2010}. Simulations are performed with different flow rate signals at the in-flow, as detailed in Fig~\ref{fig3:BreathingVsSpeaking}. The in-flow signal is perfectly periodic with a fixed cycle duration of 4.0 s for all cases reported in this paper. More details about the physical model, the numerics and the simulations are provided in the SI. 
Note that we report simulations of turbulent transient flows. Only ensemble averaging could yield results specific to each case and quantify small differences. However, we use simulations to establish trends which are common to the different cases. In the SI, the question of the reproducibility of the results and the influence of the definition of jet characteristics are notably discussed in more details.

\bibliography{referencesPNAS}

\end{document}